\documentclass{elsarticle}
\usepackage{xcolor}
\usepackage{graphicx}
\usepackage{amsmath}   
\usepackage{amsfonts}  
\usepackage{amssymb}   
\usepackage{multirow} 
\usepackage{amssymb}
\usepackage{csquotes}
\usepackage{pifont}
\usepackage{hyperref}  

\begin{document}
\begin{frontmatter}
\title{Complex Swin Transformer for Accelerating Enhanced SMWI Reconstruction}
\author{Muhammad Usman $^{1}$ and Sung-Min Gho$^{2}$}

\address{%
$^{1}$\quad Department of Anesthesiology, Perioperative and Pain Medicine, Stanford University, CA 94305, USA (usmanm@stanford.edu)}
\address{%
$^{5}$ \quad Medical R\&D Center, DeepNoid Inc., Seoul, South Korea}


\begin{abstract}
SMWI, an advanced imaging technique for detecting nigral hyperintensity in Parkinson’s Disease, is hindered by long scan times at full resolution. There is a need for efficient methods to produce high-quality SMWI from reduced k-space data. To maintain diagnostic relevance in SMWI images reconstructed from low-resolution k-space data. Complex Swin Transformer Network for super-resolving multi-echo MRI data. The method achieved SSIM of 91.16\% and MSE of 0.076 for SMWI reconstructions from 256$\times$256 k-space data, preserving diagnostic quality. This research enables high-quality SMWI imaging from reduced k-space data, accelerating scan times while preserving diagnostic detail. The approach could significantly enhance SMWI's clinical application for Parkinson’s Disease and support faster, more efficient neuroimaging workflows.

\end{abstract}
\begin{keyword}
Multi-task learning \sep semi-supervised learning \sep thyroid nodule segmentation \sep transformer \sep ultrasound images.

\end{keyword}
\end{frontmatter}

\section*{Introduction}
Susceptibility Map-Weighted Imaging (SMWI) is an advanced susceptibility-weighted imaging (SWI) technique that enhances the visibility of nigral
hyperintensity in the substantia nigra (SN), a critical feature for diagnosing Parkinson's Disease\cite{gho2014susceptibility, nam2017imaging}. Conventional SMWI protocols require extensive \textit{k}-space
data, leading to long scan times that limit clinical utility. Achieving high-quality SMWI images from under-sampled \textit{k}-space data could significantly reduce
scan times, but this often compromises diagnostic accuracy. To address this, we propose a Complex Swin Transformer Network framework designed to
enhance multi-echo data quality from highly undersampled k-space by improving resolution while preserving crucial diagnostic details for high-resolution
SMWI reconstruction.

While deep learning-based frameworks have demonstrated promising performance, their ability to capture long-range dependencies—critical for accurate enhancement remains limited \cite{farooq2025gdssa}. These limitations become more pronounced in cases with substantial variations.

Existing approaches predominantly rely on single-task learning, often overlooking complementary contextual cues such as thyroid gland structure, gland–nodule spatial interactions, and morphological priors. At the same time, the field faces a persistent shortage of large-scale annotated datasets, restricting model generalizability in real-world conditions. In contrast, recent advances in medical imaging demonstrate the benefits of incorporating auxiliary tasks and attention-driven architectures across several domains, such as mandibular canal delineation \cite{usman2022dual}, lung nodule segmentation with adaptive ROI selection \cite{usman2023deha,usman2020volumetric,usman2025multi}, brain tumor segmentation \cite{ullah2022cascade,rehman2023selective}, diabetic retinopathy segmentation \cite{ullah2023ssmd}, and broader biomedical detection challenges \cite{iqbal2023ldmres,ullah2023densely,ullah2023mtss}. These works consistently highlight how multi-scale attention, ROI adaptation, and multi-encoder feature fusion contribute to improved lesion localization and robustness.

Transformer-based architectures have further accelerated progress by enabling superior long-range context modeling, benefiting not only medical image reconstruction \cite{latif2018automating,usman2020retrospective,usman2024advancing,rehman2024biological,farooq2025anatomy} but also segmentation and detection tasks across MRI, CBCT, and CT modalities. Their ability to integrate global and local representations has proven effective in diverse clinical workflows, such as cardiomegaly assessment \cite{lee2021evaluation}, multimodal neuroimaging fusion \cite{usman2024meds,latif2020leveraging}, phonocardiographic signal analysis \cite{latif2018phonocardiographic}, and cross-lingual feature representation learning \cite{latif2018cross}. Collectively, these studies demonstrate that multi-task learning (MTL), multi-encoder architectures, and attention-rich designs consistently outperform strictly single-task CNN systems, especially in complex and low-contrast settings.

MTL in particular has shown strong potential in enhancing primary task performance by leveraging the inductive bias of complementary auxiliary tasks. In medical imaging, MTL-driven designs have yielded notable improvements in lung nodule detection \cite{usman2024meds}, metaverse-based brain-age estimation \cite{usman2024advancing}, COVID-19 classification \cite{ullah2023densely,ullah2023mtss}, and other clinical prediction tasks \cite{usman2017using,latif2018mobile}. Semi-supervised strategies also continue to play an important role in addressing data scarcity by effectively utilizing unlabeled samples for model pretraining or auxiliary task optimization \cite{latif2018automating,usman2020retrospective,usman2023deha,usman2024intelligent, usman20251575}.

\begin{figure}[h!]
  \centering
  \includegraphics[width=1.2\linewidth]{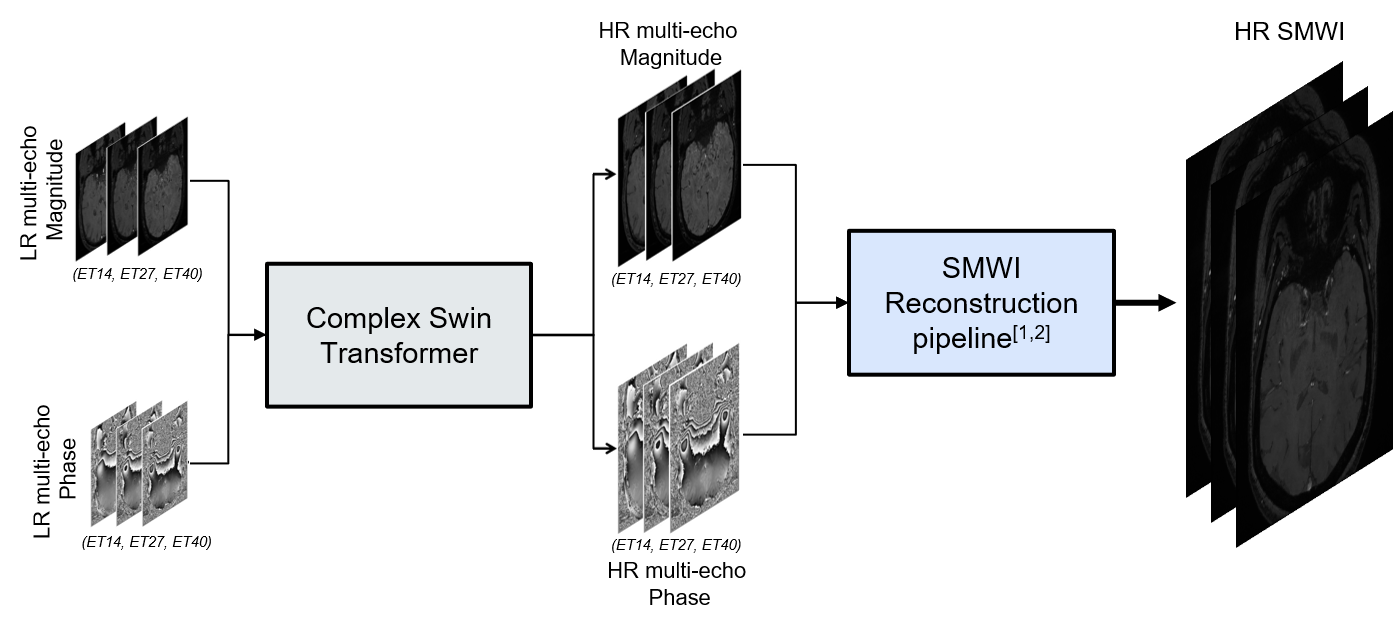}
  \caption{Illustration of our proposed pipeline to generate high resolution SMWI scan from low resolution multi-echo data.}
  \label{fig:pipeline}
\end{figure}
\section*{Proposed Method}
\begin{figure}[h!]
  \centering
  \includegraphics[width=1.2\linewidth]{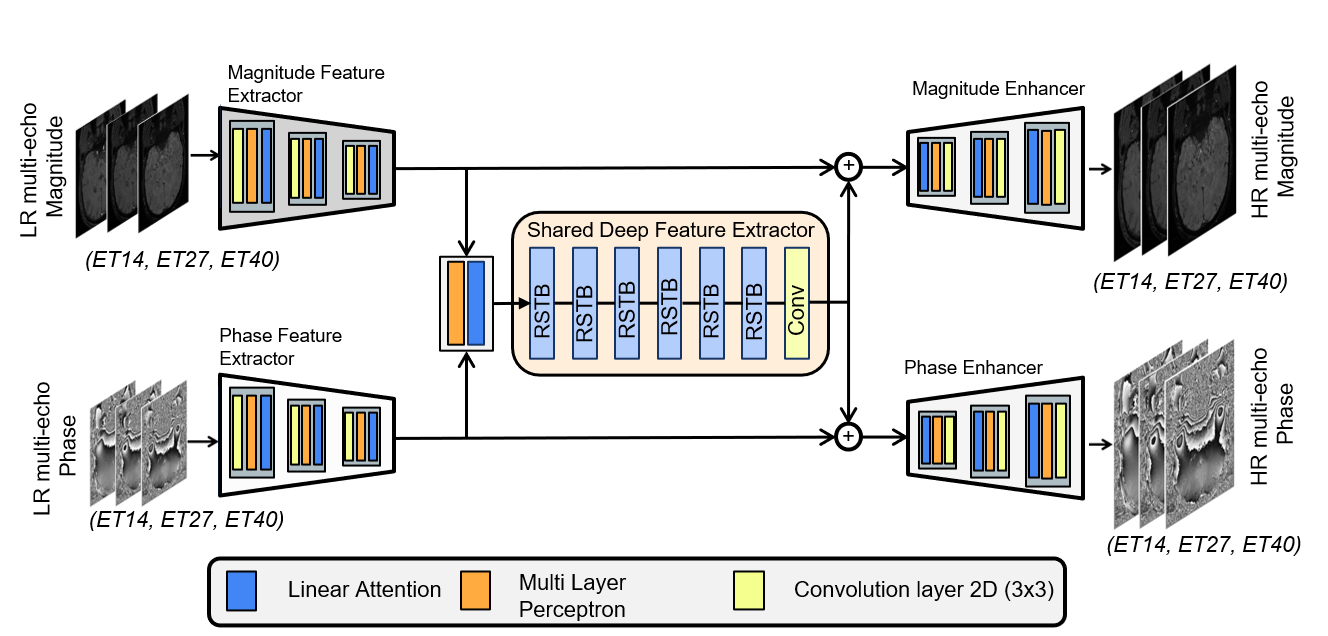}
  \caption{Illustration of our proposed Complex Swin Transformer Architecture which is designed to enhance the resolution of complex images of multi-echo scans for high resolution SMWI generation.}
  \label{fig:architecture}
\end{figure}

As illustrated in Figure~\ref{fig:pipeline}, our proposed pipeline begins with the Complex Swin Transformer Network, which enhances the resolution of low-resolution
complex data. This is followed by SMWI reconstruction to generate high-resolution SMWI images. As detailed in Figure~\ref{fig:architecture}, the Complex Swin Transformer
Network includes separate feature extractors for magnitude and phase components, a shared deep feature extractor, and two high-quality (HQ) data generators.
The complex inputs (magnitude and phase) are processed independently through dedicated feature extractors, with the resulting features then concatenated and passed
through a Shared Deep Feature Extractor comprising six Residual Swin Transformer Blocks (RSTB)~\cite{liang2021swinir} and a convolutional layer. Finally, HQ data generators
produce high-quality magnitude and phase outputs, which are further utilized for HQ SMWI reconstruction.

\section*{Dataset}
We utilized a total of 214 SMWI scans in this study, with 140 scans for training, 10 for validation, and 64 for testing. High-resolution 384$\times$384 k-space data
with three echo times (ET = 14 ms, 27 ms, and 40 ms) were acquired on a Siemens scanner to create reference images. To simulate under-sampling, we reduced the
k-space dimensions to 192$\times$192 and 256$\times$256, then converted these to the spatial domain to create low-resolution complex data for each echo.

\section*{Results}
\begin{figure}[h!]
  \centering
  \includegraphics[width=0.95\linewidth]{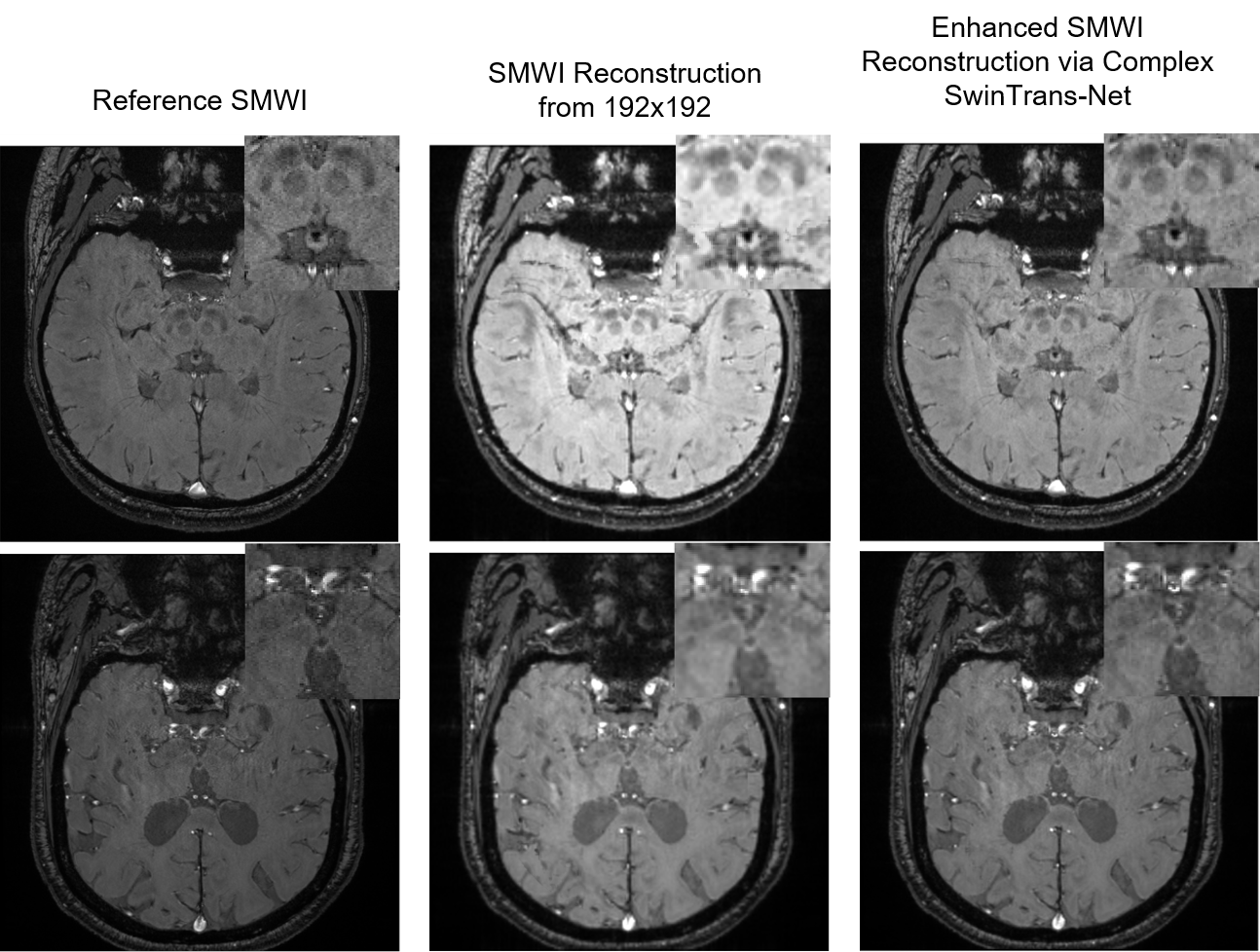}
  \caption{Demonstration of visual results obtained from low resolution data (i.e., 192$\times$192) by employing proposed complex Swin Transformer Network.}
  \label{fig:vis192}
\end{figure}

\begin{figure}[h!]
  \centering
  \includegraphics[width=0.95\linewidth]{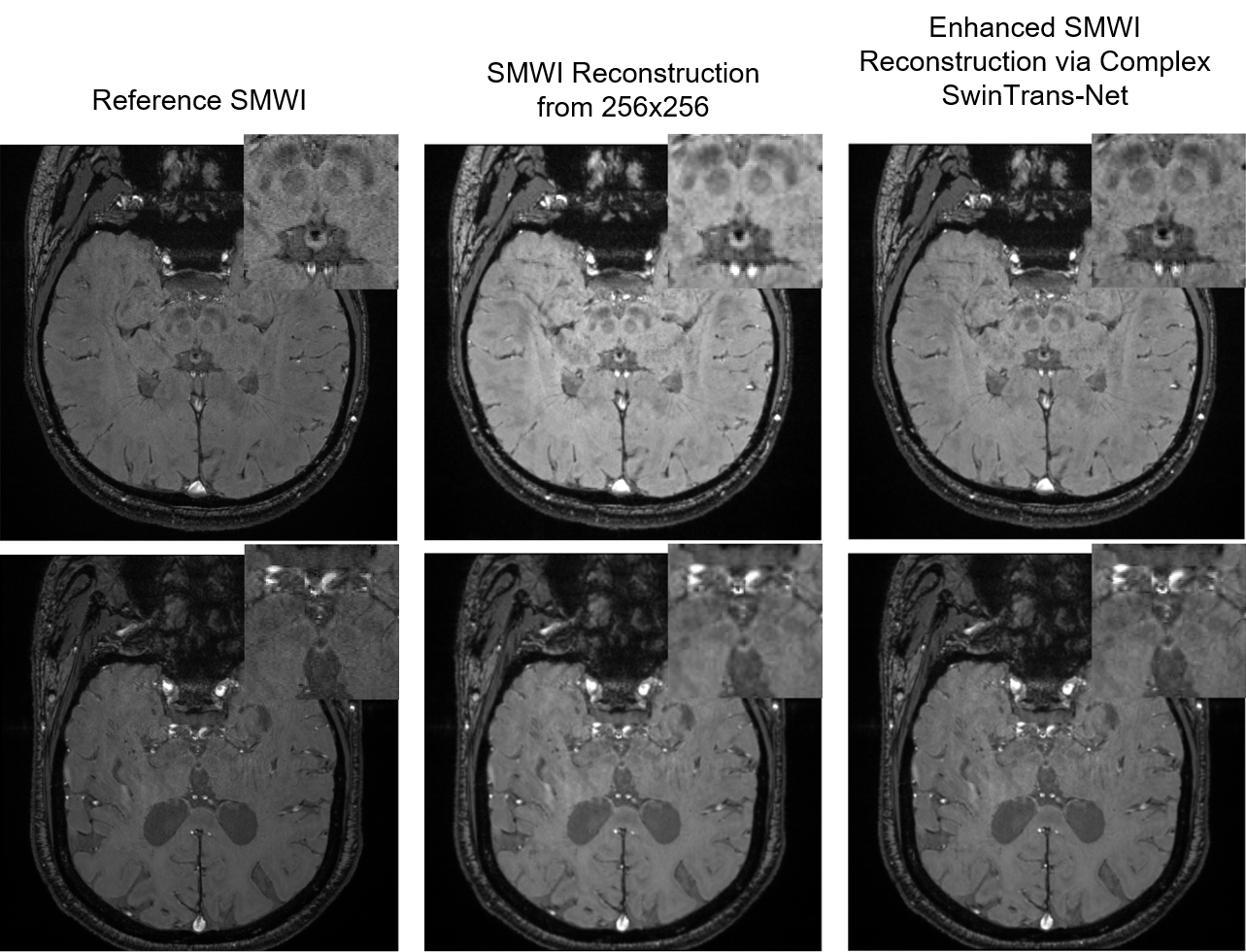}
  \caption{Demonstration of visual results obtained from low resolution data (i.e., 256$\times$256) by employing proposed complex Swin Transformer Network.}
  \label{fig:vis256}
\end{figure}

We conducted quantitative and qualitative evaluations on the 64 test scans. For k-space data of 256$\times$256, the proposed method achieved an MSE of
0.064 $\pm$ 0.050, MAE of 1.66 $\pm$ 0.635, and SSIM of 91.16 $\pm$ 2.89. For k-space data of 192$\times$192, we achieved an MSE of 0.079 $\pm$ 0.075,
MAE of 2.07 $\pm$ 1.00, and SSIM of 89.74 $\pm$ 2.53. Figures~\ref{fig:vis192} and~\ref{fig:vis256} display the qualitative results, demonstrating that our method
preserves essential diagnostic information in the reconstructed SMWI images.

\section*{Discussion and Conclusion}
This study demonstrates that the proposed Complex Swin Transformer Network effectively enhances the quality of undersampled k-space multi-echo
data, enabling high-resolution SMWI reconstruction with reduced scan times. Both quantitative and qualitative results suggest that our approach
significantly reduces the k-space data requirements, resulting in reduced scan time while maintaining diagnostic quality, making it a promising tool
for efficient clinical imaging.

\section*{Acknowledgements}
No acknowledgement found.

\bibliographystyle{elsarticle-num}
\bibliography{refs.bib}

\end{document}